\documentclass[twocolumn,showpacs,preprintnumbers,amsmath,amssymb,superscriptaddress]{revtex4-1}

\usepackage{epsf}
\usepackage{graphicx}
\usepackage{sidecap}

\usepackage{color}

\def \beq {\begin{equation}}
\def \eeq {\end{equation}}
\begin{document}

\title{{Tunability of the topological nodal-line semimetal phase in ZrSiX-type materials}}





\author{M.~Mofazzel~Hosen}\affiliation {Department of Physics, University of Central Florida, Orlando, Florida 32816, USA}

\author{Klauss~Dimitri}\affiliation {Department of Physics, University of Central Florida, Orlando, Florida 32816, USA}

\author{Ilya~Belopolski}
\affiliation {Joseph Henry Laboratory and Department of Physics, Princeton University, Princeton, New Jersey 08544, USA}

\author{Pablo Maldonado}
\affiliation {Department of Physics and Astronomy, Uppsala University, P. O. Box 516, S-75120 Uppsala, Sweden}


\author{Raman Sankar} \affiliation{Institute of Physics, Academia Sinica, Taipei 10617, Taiwan} 
\affiliation{Center for Condensed Matter Sciences, National Taiwan University, Taipei 10617, Taiwan}







\author{Nagendra Dhakal}
\affiliation {Department of Physics, University of Central Florida, Orlando, Florida 32816, USA}

\author{Gyanendra Dhakal} 
\affiliation {Department of Physics, University of Central Florida, Orlando, Florida 32816, USA}

\author{Taiason Cole}
\affiliation {Department of Physics, University of Central Florida, Orlando, Florida 32816, USA}



\author{Peter M. Oppeneer}
\affiliation {Department of Physics and Astronomy, Uppsala University, P. O. Box 516, S-75120 Uppsala, Sweden}

\author{Dariusz Kaczorowski}
\affiliation {Institute of Low Temperature and Structure Research, Polish Academy of Sciences,
50-950 Wroclaw, Poland}

\author{Fangcheng Chou} \affiliation{Center for Condensed Matter Sciences, National Taiwan University, Taipei 10617, Taiwan}

\author{M.~Zahid~Hasan}
\affiliation {Joseph Henry Laboratory and Department of Physics,
Princeton University, Princeton, New Jersey 08544, USA}

\author{Tomasz~Durakiewicz}
\affiliation {Condensed Matter and Magnet Science Group, Los Alamos National Laboratory, Los Alamos, NM 87545, USA} 
\affiliation {Institute of Physics, Maria Curie - Sklodowska University, 20-031 Lublin, Poland}

\author{Madhab~Neupane}
\affiliation {Department of Physics, University of Central Florida, Orlando, Florida 32816, USA}

\date{18 June, 2013}
\begin{abstract}

{The discovery of a topological nodal-line (TNL) semimetal phase in ZrSiS has invigorated the study of other members of this family. Here, we present a comparative electronic structure study of ZrSiX (where X = S, Se, Te) using angle-resolved photoemission spectroscopy (ARPES) and first-principles calculations. Our ARPES studies show that the overall electronic structure of ZrSiX materials comprises of the diamond-shaped Fermi pocket, the nearly elliptical-shaped Fermi pocket, and a small electron pocket encircling the zone center ($\Gamma$) point, the M point, and the X point of the Brillouin zone, respectively. We also observe a small Fermi surface pocket along the M-$\Gamma$-M direction in ZrSiTe, which is absent in both ZrSiS and ZrSiSe. Furthermore, our theoretical studies show a transition from nodal-line to nodeless gapped phase by tuning the chalcogenide from S to Te in these material systems. Our findings provide direct evidence for the tunability of the TNL phase in ZrSiX material systems by adjusting the spin-orbit coupling (SOC) strength via the X anion.}





\end{abstract}
\date{\today}
\maketitle

  

Topological insulators (TIs) are bulk insulating systems with gapless robust surface states originating from a non-trivial electronic band structure \cite{Hasan, SCZhang, Hasan_review_2, Xia, Neupane_4}. These systems are on the frontier of research interests of modern condensed matter physics. Recently semimetals such as the Dirac semimetal, Weyl semimetal, and nodal-line semimetal have been reported to also exhibit this novel non-trivial topological state \cite{Neupane, Neupane_2,wang, wang1, Nagaosa, Young_Kane, Dai, Dai_LiFeAs, Suyang_Science, Hasan_2, Hong_Ding, TaAs_theory_1, TaAs_theory, node_0, node_1, node_2, node_3, node_4, Neupane_5, Schoop, Ding, tranp1, Xi_Dai, Suyang, Chen, Weyl, Cava1}. The Dirac and Weyl semimetals possess symmetry protected zero dimensional (0D) band touching near the Fermi level \cite{Neupane, Neupane_2,Cava1, wang, wang1, Nagaosa, Young_Kane, Dai, Dai_LiFeAs, Suyang_Science, Hasan_2, Hong_Ding, TaAs_theory_1, TaAs_theory}. These states have been experimentally demonstrated in many materials such as  Cd$_3$As$_2$ \cite{Dai, Neupane_2, Cava1, wang, wang1}, Na$_3$Bi \cite{Xi_Dai, Suyang, Chen, Weyl}, TaAs \cite{TaAs_theory_1, TaAs_theory, Suyang_Science, Hong_Ding}, NdSb \cite{NdSb}, WTe$_2$ \cite{WT, wang2, Ber}, etc. Their low-energy bulk band structure contains linearly dispersive bands, leading to unusual exotic transport behaviors like extremely high bulk carrier mobility and extreme magnetoresistance as a consequence of the chiral anomaly \cite{Cava, Chiral1, Chiral2, wang3}. Moreover, nodal-line semimetals show extended line like one dimensional (1D) band touching loops or lines in momentum (k) space. These 1D loops or lines need extra symmetries to protect them aside from the translational symmetry, and show many anomalous properties in transport measurements \cite{node_0, node_1, node_2, node_3, node_4, Dai_LiFeAs, Neupane_5, Schoop, Ding, tranp1}. To date, several materials such as PbTaSe$_2$ \cite{PTS}, PtSn$_4$ \cite{PT}, and ZrSiS \cite{Neupane_5, Schoop} have been experimentally reported to host this novel state.
  
The recently reported ZrSiS is an interesting material because it exhibits multiple Dirac cones as well as the nodal-line semimetal phase \cite{Neupane_5, Schoop}. Importantly, the Dirac bands in this material show a linear dispersion behavior over a large energy range  of up to 2 eV. Furthermore, a single layer of this material is predicted to host the 2D TI electronic structure \cite{Dai_LiFeAs, Young_Kane}. The nodal-line semimetal phase in ZrSiS was reported assuming negligible SOC strength \cite{Neupane_5, Schoop}. Hence, by tuning the chalcogenide from S to Te, while maintaining the isoelectronic properties but by increasing the atomic number (Z) of the chalcogen ion, one can play with the SOC strength to investigate the possible tunability of the nodal phase in this material system. Recently, the existence of the topological nodal-line semimetal phase in ZrSiSe and ZrSiTe as well as the 3D to 2D structural transition in ZrSiTe have been suggested \cite{new_ZST}. The two compounds ZrSiSe and ZrSiTe have a layered crystal structure similar to ZrSiS but with a different SOC strength. However, the details of the electronic structure as well as the effect of SOC have not been reported in these material systems using momentum-resolved spectroscopic probe and first-principles band structure calculations.

\begin{figure*}
	\centering
	\includegraphics[width=17.00cm]{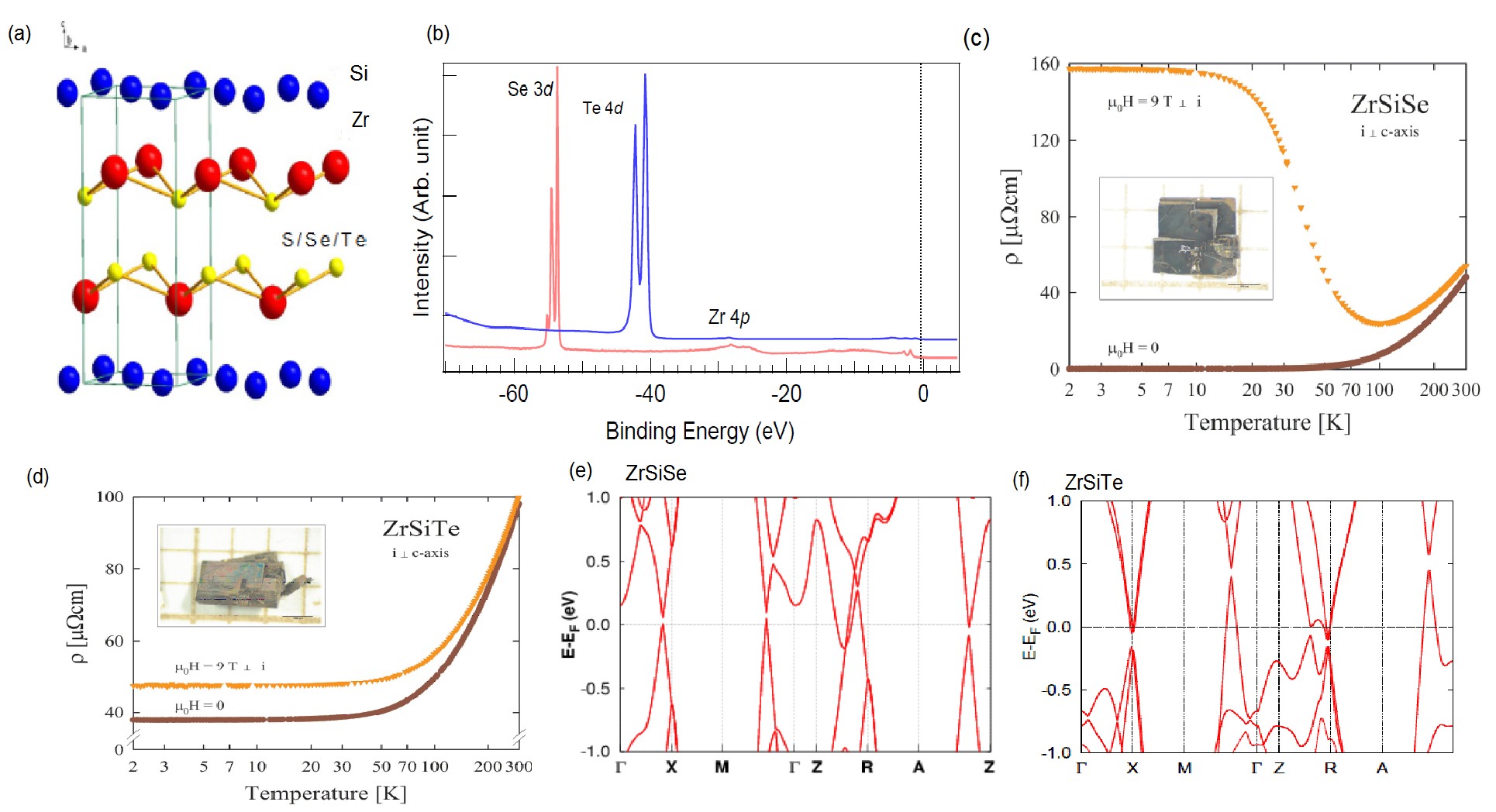}
	\caption{{Crystal structure and sample characterization}. (a) Tetragonal crystal structure of ZrSiX  (X = S, Se, Te). (b) Core level spectroscopic measurements for ZrSiSe (red curve) and ZrSiTe (blue curve). (c)-(d) Sample picture and temperature dependent resistivity plot  with current along ab direction of ZrSiSe and ZrSiTe, respectively. (e)-(f) Calculated band structures of ZrSiSe and ZrSiTe along high symmetry directions, with symmetry points indicated.} 
	
\end{figure*}

Here, we present a comparative electronic structure study of ZrSiX (where X = S, Se, Te) using angle-resolved photoemission spectroscopy (ARPES) and first-principles calculations to reveal the effect of SOC on their band structures. Our ARPES measurements show that the overall electronic structures are the same for ZrSiS, ZrSiSe and ZrSiTe, but the details of their electronic structure show subtle but important differences. We observe a diamond-shaped Fermi surface around the zone center ($\Gamma$) point, a nearly elliptical-shaped Fermi pocket around the M point, and a small electronic pocket at the corner (X) point of the Brillouin zone (BZ) in ZrSiSe and ZrSiTe. However, in ZrSiTe an additional small pocket is observed in the M-$\Gamma$-M momentum space direction. Furthermore, we observe linearly dispersive surface states in both materials around the X point of the BZ. These experimental observations are further supported by our first-principles band structure calculations. Our theoretical band structure calculation shows a negligible gap at ZrSiSe and an approximately 60 meV gap in ZrSiTe around the M-$\Gamma$-M (nodal-line) direction in the BZ. This gap opening is attributed to the higher SOC effect in ZrSiTe. Our findings provide a platform to observe the tunable nodal-line semimetal phase in the ZrSiX material system.
\begin{figure*}
	\centering
	\includegraphics[width=17.00cm]{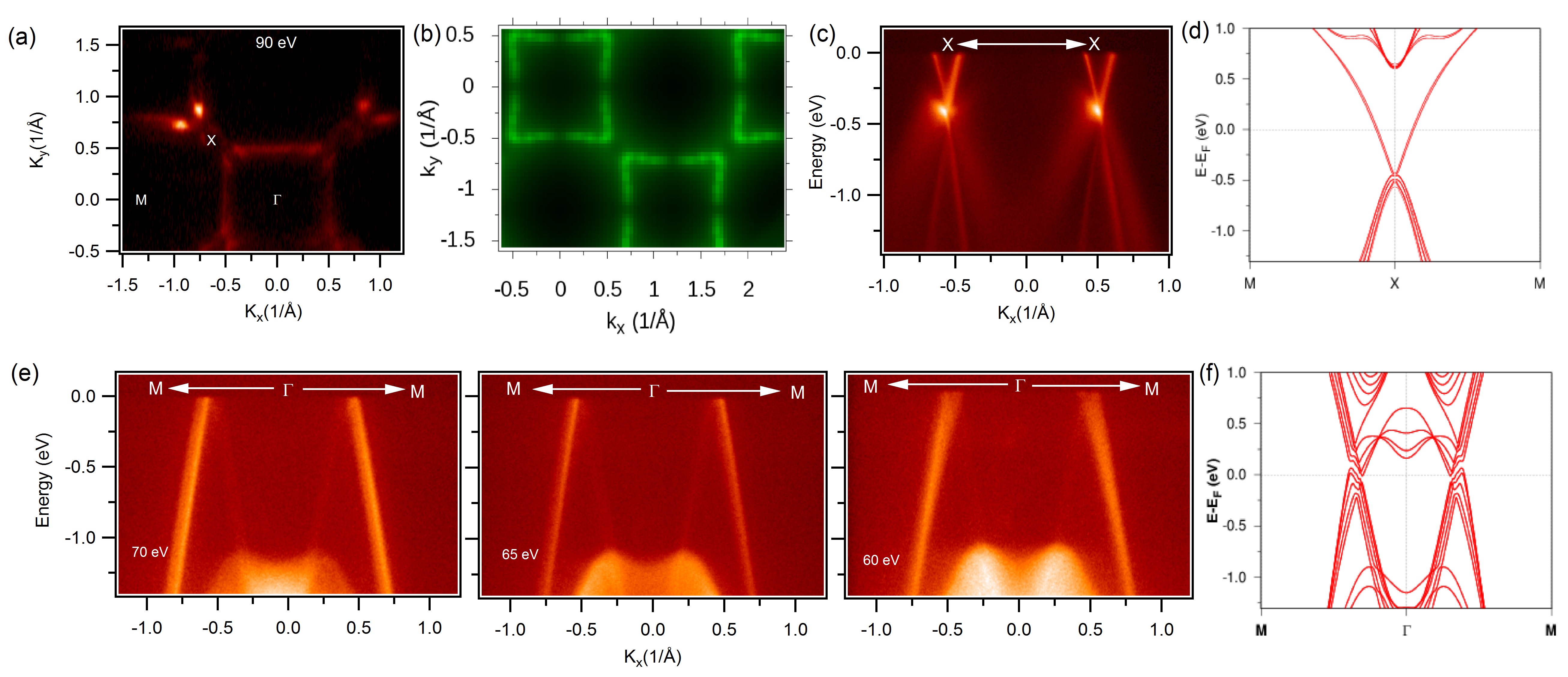}
	\caption{{Electronic structure of ZrSiSe.} (a) Fermi surface map measured with a photon energy of 90 eV. High symmetry points are noted. (b) Calculated Fermi surface cross-section. (c) Dispersion map along the X-X direction. (d) Band dispersion obtained from slab calculations along the M-X-M direction. (e) Photon energy dependent ARPES dispersion along the M-$\Gamma$-M direction. The photon energies are also marked in the figures. (f) Calculated bands along the M-$\Gamma$-M direction. ARPES measurements presented in this figure were performed at ALS BL 4.0.1 at a temperature of 18 K.}
\end{figure*}

The single crystals of ZrSiSe and ZrSiTe were grown by the vapor transport method as described elsewhere \cite{Material, growth}. Their chemical composition was checked by energy-dispersive X-ray analysis using a FEI scanning electron microscope equipped with an EDAX Genesis XM4 spectrometer. The results indicated homogeneous single-phase materials with their stoichiometry close to equiatomic. Single crystal X-ray diffraction (XRD) experiments were performed on a Kuma-Diffraction KM4 four-circle diffractometer equipped with a CCD camera using Mo K$\alpha$ radiation. Rietveld refinements of the XRD data yielded the tetragonal space group P4/nmm with the lattice parameters close to those reported in the literature \cite{Material}. Electrical resistivity measurements were carried out within the temperature range 2-300 K and in applied magnetic fields up to 9 T using a conventional four-point ac technique implemented in a Quantum Design PPMS platform. The electrical contacts to as-grown crystals were made using a silver epoxy paste. The electrical resistivity of ZrSiSe and ZrSiTe was measured with electrical current flowing within the basal plane of their tetragonal unit cells. Synchrotron-based ARPES measurements of the electronic structure  were performed at the SIS-HRPES end station of the Swiss Light Source (SLS) and Advanced Light Source (ALS) BL 4.0.1 with Scienta R4000 and R8000 hemispherical electron analyzers. The energy resolution was set to be better than 20 meV  and the angular resolution was set to be better than  0.2$^{\circ}$ for these measurements. The electronic structure calculations were carried out using the Vienna ab initio simulation package (VASP) \cite{Kress_1, VASP}, with the generalized gradient approximation (GGA) as the density functional theory exchange-correlation functional  \cite{ GGA_1}. Projector augmented-wave pseudopotential \cite{Blochl_1} were used with an energy cutoff of 500 eV for the plane-wave basis, which was  sufficient to converge the total energy for a given k-point sampling. In order to simulate surface effects, we used a 1 $\times$ 1 $\times$ 5 supercell for the (010) surface, with a vacuum thickness larger than 19  \AA. The Brillouin zone integrations were performed on a special \textit{k}-point mesh generated by a 25 $\times$  25 $\times$  25 and a 40 $\times$  40 $\times$  1 $\Gamma$-centered Monkhorst Pack \textit{k}-point grid for the bulk and surface calculations, respectively. The SOC was included self-consistently in the electronic structure calculations. 



We start our discussion by presenting the crystal structures of ZrSiSe and ZrSiTe. Similar to the ZrSiS structure they also crystallize in a PbFCl-type crystal structure with space group $P4/nmm$ \cite{Material}. Zr layers are separated by two neighboring X atoms layers, and sandwiched between a Si square net [see Fig. 1(a)]. The bond between Zr-X is relatively weak, and it  provides a natural cleavage plane between two neighboring Zr-Se or Zr-Te terminals. The crystal easily cleaves along the (001) surface. Fig. 1(b) shows the core level measurements of ZrSiSe and ZrSiTe. The red curve represents the spectroscopic core level measurement of ZrSiSe and blue curve represents the core level measurement of ZrSiTe. We observe a peak of Zr 4$p$ around 27 eV for both materials, and sharp peaks of Te 4$d$$_{5/2}$ ($\sim$ 40.7 eV), Te 4$d$$_{3/2}$ ($\sim$ 42 eV) of ZrSiTe, Se 3$d$ ($\sim$ 55 eV) of ZrSiSe. It indicates that the sample used for our measurements is of good quality. Fig. 1(c) and Fig. 1(d) display the temperature variations of the electrical resistivity of single-crystalline ZrSiSe and ZrSiTe, respectively, measured with electrical current flowing within the basal plane of their tetragonal unit cells (i $\perp$ c-axis). In a zero magnetic field, both compounds exhibit metallic conductivity. For ZrSiSe, the ratio of the room-temperature resistivity to the residual resistivity of 0.3 $\mu$$\Omega$cm observed at 2 K was about 160, proving the excellent quality of the single crystal studied. In the case of ZrSiTe, these two quantities were 38 $\mu$$\Omega$cm and 2.6, respectively, which might signal possible atomic disorder in the crystal lattice. Another plausible explanation of this result is some oxidation of the surface of the measured specimen that resulted in worse electrical contacts (we observed that crystals of the telluride, in contrast to those of ZrSiS and ZrSiSe, are fairly unstable against air and moisture). In a magnetic field applied perpendicular to the electrical current, the electrical resistivity of ZrSiSe changes in a previously reported manner for topologically nontrivial semimetals, like WTe$_2$ \cite{WT, Ber}, LaSb \cite{LaSb}, TaAs$_2$ \cite{TaAs2}.

\begin{figure*}
	\centering
	\includegraphics[width=17.0cm]{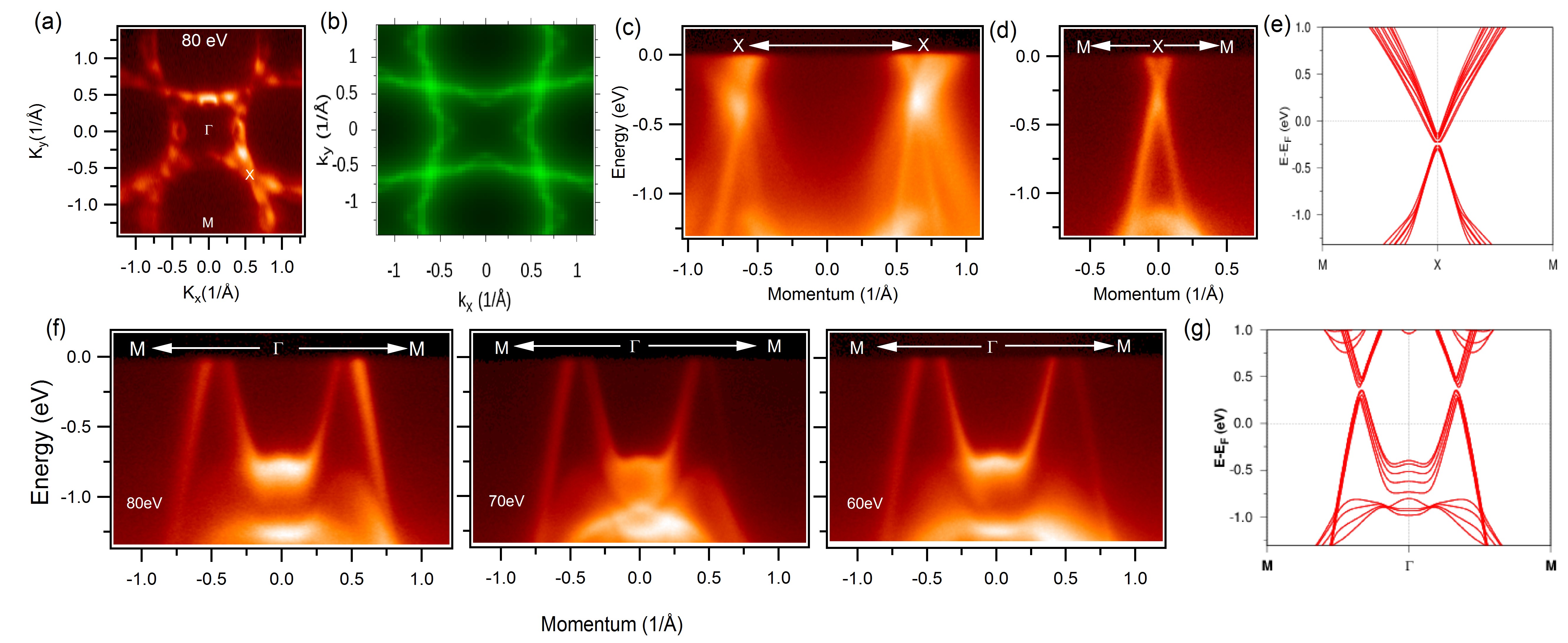}
	\caption{{Electronic structure of ZrSiTe.}
		(a) Fermi surface measured with a photon energy of 80 eV. High symmetry points are noted in the plot. (b) Calculated Fermi surface map. (c) Dispersion map along the X-X direction measured with photon energy of 60 eV. (d) Measured dispersion map along the M-X-M direction. (e) Calculated bands along the M-X-M direction. (f) Photon energy dependent dispersion maps along the M-$\Gamma$-M direction. Photon energies are mentioned in the plots. (g) Calculation showing the dispersion map along the M-$\Gamma$-M high-symmetry direction. These data were collected at the SIS-HRPES end station at SLS at a temperature of 18 K.}
\end{figure*}

 The characteristic upturn in $\rho$(T) followed by a plateau at low temperatures can be attributed to magnetic-field-driven changes in charge carrier densities and mobilities in a system close to nearly perfect electron-hole compensation \cite{compen}. An alternative explanation for the observed behavior is a magnetic-field-induced metal-insulator transition in Dirac systems \cite{WT}. In a field of 9 T, the transverse (H $\perp$ i) magnetoresistance (MR) of ZrSiSe, defined as MR = [$\rho$(T,H) - $\rho$(T,0]/$\rho$(T,0), reaches at 2 K a giant value of about 48,000$\%$. With increasing temperature, it systematically decreases down to 13$\%$ at 300 K. In contrast to the selenide, ZrSiTe exhibits ordinary magnetotransport, characterized by positive MR values of 26$\%$ at 2 K and 3$\%$ at 300 K, in a field of 9 T. This behavior hints at prevalence of one type of carrier which determines the electrical conductivity of the system. Fig. 1(e) shows the band structure of ZrSiSe calculated with including the SOC, along various high symmetry directions in the BZ. A negligible gap is found at the M-$\Gamma$-M direction. Furthermore, compared to the previously reported band calculation of ZrSiS \cite{Neupane_5, Schoop}, bands are shifted towards the chemical potential at various high symmetry points. Fig. 1(f) shows the calculated band structure of ZrSiTe. Compare to ZrSiSe (see Fig. 1e) some bands are shifted upward and others downward around the various high symmetry points. 


In order to determine the details of the electronic structure of ZrSiSe and ZrSiTe, we present our systematic surface electronic structure studies in Figs. 2 and 3. Fig. 2(a) shows the ARPES measured Fermi surface map at a photon energy of 90 eV at a temperature of 18 K. We observe a diamond shape Fermi surface around the $\Gamma$ points, a nearly elliptical-shaped Fermi pocket around the M point and a small circular electron pocket at the X point of the BZ, which is in agreement with the previously established electronic structure of ZrSiS \cite{Neupane_5, Schoop}. The high symmetry points are also denoted in the figure. Fig. 2(b) shows the calculated Fermi surface. The calculated Fermi surface shows good agreement with the experimental observation. Fig. 2(c) shows the dispersion map along the X-X direction measured with a photon energy of 65 eV. The Dirac point is observed nearly 400 meV below the chemical potential.
One possible reason behind this brightness could be the matrix element effect. Fig. 2(d) shows the calculated bands along the M-X-M direction. The photon energy dependent ARPES dispersion along the high symmetry M-$\Gamma$-M direction is presented in Fig. 2(e). These states originate from the nodal-line state, and show excellent agreement with our calculation in Fig. 2(f). The nodal-line points from our measured dispersion maps are not visible along this direction (M-$\Gamma$-M) as these are located above the observed chemical potential. We note that the linearly dispersive band energy range is larger than 1 eV below the chemical potential.

In Fig. 3 we present our electronic structure study of ZrSiTe. Fig. 3(a) presents the Fermi surface map of ZrSiTe measured with a photon energy of 80 eV and a temperature of 18 K. The overall electronic structure of ZiSiTe is similar to that of ZrSiS and ZrSiSe. Interestingly, additional small pockets are observed along the M-$\Gamma$-M directions of the BZ. Fig. 3(b) shows the calculated Fermi surface, which is in good agreement with the experimental Fermi surface. The measured ARPES dispersion along the X-X direction is presented in Fig. 3(c). Fig. 3(d) shows the measured dispersion along the M-X-M direction. The Dirac point is observed around 300 meV below the chemical potential on both plots. Calculated bands along the M-X-M direction are shown in Fig. 3(e).  We note that, the location of the Dirac point about the X point in ZrSiSe is 400 meV below the chemical potential (see Fig. 2c) whereas that one for ZrSiTe is about 300 meV (see Fig. 3c and d). Fig. 3(f) presents the measured photon energy dependent dispersion maps along the M-$\Gamma$-M direction.  Experimentally measured dispersion maps agree well with the calculated band dispersion along the M-$\Gamma$-M direction as shown in Fig. 3(g). We observe that the linearly dispersive band along this direction extends as far as 1 eV below the chemical potential. 


To compare the electronic structures, we show the constant energy contour plots and Fermi surface maps of ZrSiX where X = S, Se, Te, in Fig. 4. Our systematic studies show that ZrSiSe has a similar electronic structure to ZrSiS \cite{Neupane_5, Schoop} but ZrSiTe possesses some different features such as a small pocket along the M-$\Gamma$-M direction. Figure 4(a) shows a plot of the Fermi surface map of ZrSiSe (left) and several constant energy contours measured with various binding energies as noted in the plots, which are measured at a photon energy of 80 eV and a temperature of around 18 K. Experimentally, we observe a diamond-shaped Fermi surface [Fig. 4(a), left]. The Dirac point is observed at the X point of the Fermi surface at around 400 meV below the Fermi level (FL) [see Fig. 4(a), right]. Moving towards higher binding energies the diamond-shaped Fermi surface with gapped out Dirac points becomes disconnected, and the structure resembles a diamond within a diamond as seen in Fig. 4(a) (right). The inner diamond shape is clearly disconnected along the M-$\Gamma$-M directions. Fig. 4(b) shows the Fermi surface map (left) and constant energy contour plots (right) of ZrSiTe at a photon energy of 80 eV. A similar diamond-like shape is observed with extra small pockets, and a Dirac point-like state is observed at the X points nearly 300 meV below the chemical potential. Moreover, we notice that the half circular like pocket size is enlarging with the increase of the binding energy indicating its hole-like nature. Furthermore, we performed measurements of Fermi surface mapping at various photon energies (see Supplementary Information), which show that hole-like pockets do not disperse with photon energy, indicating their 2D nature \cite{new_ZST}. Fig. 4(c)-(e) present the Fermi surface maps of ZrSiS (90 eV), ZrSiSe (80 eV), and ZrSiTe (90 eV) using ARPES. Approaching ZrSiTe from ZrSiS, we observe a similar kind of diamond shape Fermi surface with some extra features added in ZrSiTe. The same subtle differences are also observed in our band structure calculations (See also Supplementary Information). Furthermore, we observe a gapped out nodal line phase in ZrSiTe compared to negligible gap observed in ZrSiSe. 
 
 \begin{figure}
 	\centering
 	\includegraphics[width=9.00cm]{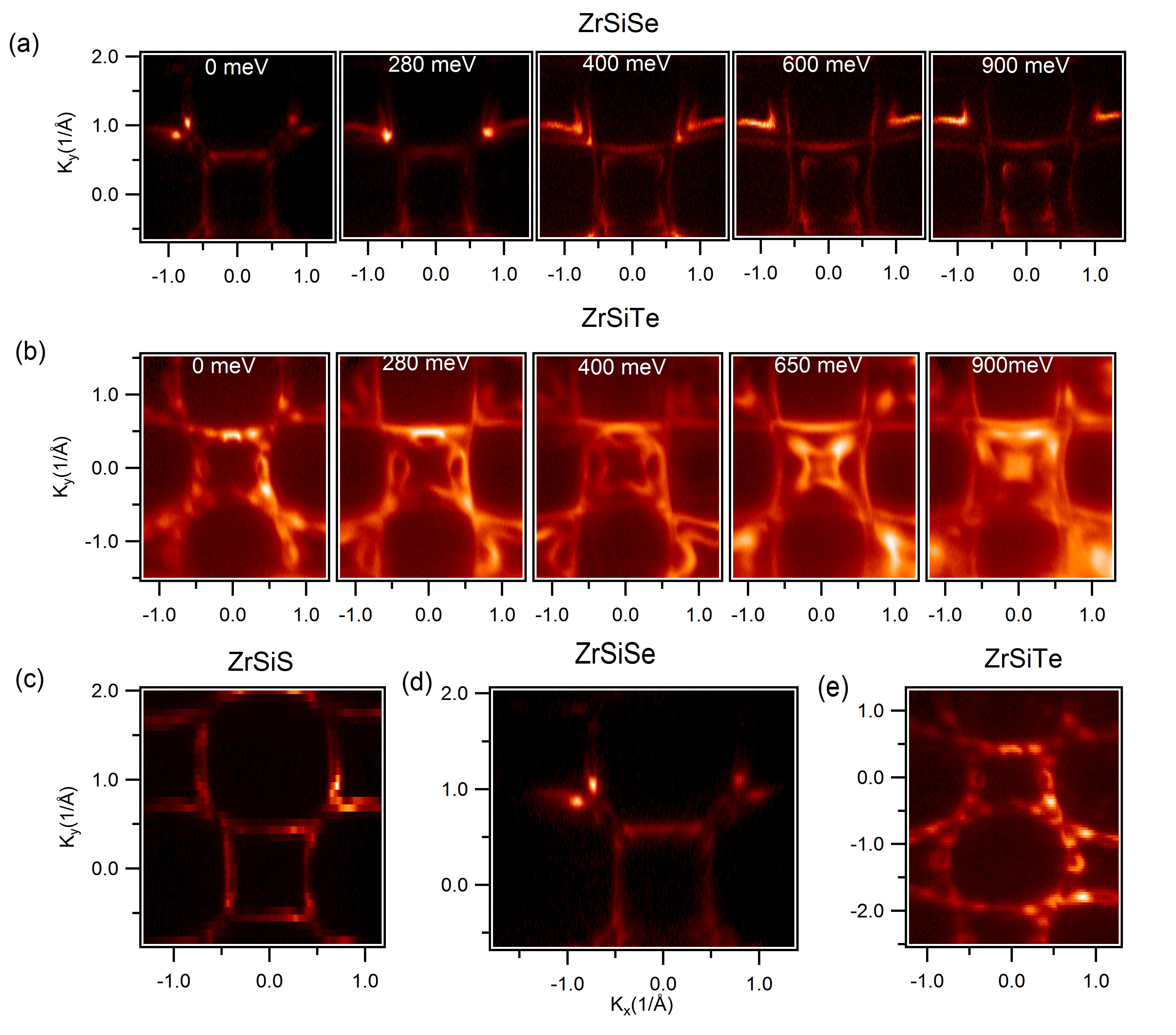}
 	\caption{{Comparison of the electronic structures of ZrSiS, ZrSiSe, and ZrSiTe}.
 		Fermi surface (0meV) and constant energy contour maps of ZrSiSe (a), and ZrSiTe (b). The binding energy values are given in the plots. (c)-(e) ARPES measured Fermi surface for ZrSiS, ZrSiSe, and ZrSiTe, respectively.}
 \end{figure}
  
Now, we discuss some observations of our experimental data and corresponding calculations. First, the overall electronic structures of ZrSiX (where X = S, Se, Te) are similar, however moving to the higher atomic number of the chalcogen (ZrSiSe $\rightarrow$ ZrSiTe) small hole-like pockets along the M-$\Gamma$-M direction appear in the Fermi surface map.  Second, our calculations show a negligible gap along the nodal direction (M-$\Gamma$-M) for ZrSiS and ZrSiSe (see Fig. 2(f)), but a gap of about 60 meV is observed in ZrSiTe (see Fig. 3(g)). The calculations show the presence of gaps along the nodal direction, but these gaps are not accessible by standard ARPES measurements because the native chemical potential resides in the valence band. Third, our result shows that the Dirac cone gap is present around the X points on these material systems. Based on the proposal of nonsymmorphic symmetry by Young and Kane, we believe this is due to the protection of these states by non-symmorphic symmetry, and at this point the Dirac cone state is not affected much by the SOC strength. Fourth, moving to the higher atomic number of the chalcogen (ZrSiS $\rightarrow$ ZrSiTe), we found small pockets at the Fermi surface, which are likely to be 2D in nature (see Fig. 3a, 4b, 4e and Supplementary Figures). Fifth, based on our calculations, the ZrSiX material system can be tuned by increasing the atomic number of the chalcogen in order to show a tunable topological nodal-line semimetal phase. 
Sixth, in the absence of evidence for chiral anomaly, we propose that the difference in response to a magnetic field among the members of the series comes from the balance in electron and hole concentrations. Here, ZrSiS and ZrSiSe are nearly compensated semimetals, while ZrSiTe is dominated by one type of carrier, as manifested by a shift of the native chemical potential. Nodal direction states in ZrSiS and ZrSiSe are located in the vicinity of the chemical potential, and scattering is suppressed in zero field. In the  case of ZrSiTe, the nodal phase states are already gapped out at zero field, so the moderate magnetic field hardly affects charge concentrations and charge mobilities, and magnetotransport becomes trivial.

In conclusion, we have performed a comparative electronic structure study of the ZrSiX system using ARPES and first-principles calculations. Our experimental data reveal the overall similarities among electronic structures of ZrSiX with an additional feature of a small hole pocket in ZrSiTe, which is complemented by our first-principles calculations. 
Importantly, our theoretical study reveals the exciting possibility of the continuous transition of the topological nodal-line semimetal fermion phase to the nodeless gapped out semimetal phase in these systems by tuning the chalcogen element of these materials. The changing SOC effect can be the reason behind this gap opening when the system moves from ZrSiS to ZrSiTe. 
Our studies reveal the systematic electronic structure of ZrSiX, and establish a new material system with an SOC tunable nodal-line phase.
 \bigskip
 \bigskip
 
 M.N. is supported by the start-up fund from the University of Central Florida.
 T.D. is supported by NSF IR/D program. 
 I. B. acknowledges the support of the NSF GRFP.
 D.K. was supported by the National Science Centre (Poland) under research grant 2015/18/A/ST3/00057.
 Work at Princeton University is supported by the Emergent Phenomena in Quantum Systems Initiative of the Gordon and Betty Moore Foundation under Grant No. GBMF4547 (M.Z.H.) and by the National Science Foundation, Division of Materials Research, under Grants No. NSF-DMR-1507585 and No. NSF-DMR-1006492. P.M. and P.M.O. acknowledge support from the Swedish Research Council (VR), the K. and A. Wallenberg Foundation and the Swedish National Infrastructure for Computing (SNIC). We thank Plumb Nicholas Clark for beamline assistance at the SLS, PSI. We also thank Sung-Kwan Mo and Jonathan Denlinger for beamline assistance at the LBNL.

\end{document}